\documentclass[letterpaper,conference,final]{IEEEtran}

\usepackage[binary-units=true]{siunitx}
\DeclareSIUnit \dBm {dBm}
\DeclareSIUnit \dB {dB} 
\DeclareSIUnit \dBi {dBi} 
\DeclareSIUnit \Kbps {Kbps}
\DeclareSIUnit \Mbps {Mbps}
\DeclareSIUnit \Gbps {Gbps}
\DeclareSIUnit \kBps {kBps}
\DeclareSIUnit \MBps {MBps}
\DeclareSIUnit \GBps {GBps}

\usepackage{cite}      
\usepackage[caption=false, font=footnotesize]{subfig}
\usepackage{graphicx}
\usepackage{url}       
\usepackage{amsfonts}
\usepackage{amsmath}
\usepackage{times}
\usepackage{algpseudocode}
\usepackage{fourier}
\usepackage{algorithm}
\usepackage{enumerate}
\usepackage{multirow}
\usepackage{tikz}
\usepackage{array}
\usepackage{gensymb}
\usepackage{nicefrac}
\newcolumntype{P}[1]{>{\centering\arraybackslash}p{#1}}
\newcolumntype{M}[1]{>{\centering\arraybackslash}m{#1}}

\bstctlcite{IEEEexample:BSTcontrol}
\newcolumntype{C}[1]{>{\centering\let\newline\\\arraybackslash\hspace{0pt}}m{#1}}

\begin{document}
\title{Agile Calibration Process of Full-Stack Simulation Frameworks for V2X Communications}
\author{\IEEEauthorblockN{Ioannis Mavromatis, Andrea Tassi, Robert J. Piechocki, and Andrew Nix}
  \IEEEauthorblockA{Department of Electrical and Electronic Engineering, University of Bristol, UK \\ Emails: \{Ioan.Mavromatis, A.Tassi, R.J.Piechocki, Andy.Nix\}@bristol.ac.uk}
}

\maketitle

\begin{abstract}
Computer simulations and real-world car trials are essential to investigate the performance of Vehicle-to-Everything (V2X) networks. However, simulations are imperfect models of the physical reality and can be trusted only when they indicate agreement with the real-world. On the other hand, trials lack reproducibility and are subject to uncertainties and errors. In this paper, we will illustrate a case study where the interrelationship between trials, simulation, and the reality-of-interest is presented. Results are then compared in a holistic fashion. Our study will describe the procedure followed to macroscopically calibrate a full-stack network simulator to conduct high-fidelity full-stack computer simulations.
\end{abstract}


\section{Introduction}
Connected Autonomous Vehicles (CAVs) are enhanced daily with new autonomous features, leading gradually to fully autonomous means-of-transports~\cite{fully_autonomy}. Being part of the ecosystem of Intelligent Transportation Systems (ITSs), CAVs will require an agile interconnecting framework~\cite{broadband,TassiTVT}, providing a constant service and optimal system behavior. To optimize and further enhance the performance of this framework, initial experimental evaluation of real-world trials and simulation results is required.

In this work, we aim to establish a connection between simulated and trial-based results for a vehicular network. Introducing the procedure followed, we will describe how inconsistencies during the experiments are identified and excluded from the evaluation. We will later present a calibration framework for the fine-tuning of the imperfect simulation results to enhance their behavior and achieve high-fidelity ``real-world'' results.

Simulations are approximated models of the physical world. However, they are easily and inexpensively conducted using an appropriate network simulator achieving near-perfect results. For example, the number of vehicles within a network can be easily scaled up to increase the network congestion (as in~\cite{dense_network}). Furthermore, they offer a high degree of flexibility. Using different configurations and isolating particular parameters, we can examine the behavior of a system under specific conditions. This is the case of~\cite{adaptive_beaconing} where the impact of different beacon intervals has been investigated with respect to the end-to-end delay, for different road networks and city-wide scenarios.

On the contrary, real-world trials are based on a ``perfect'' model. For example, authors in~\cite{experiments} experimentally analyzed the performance of a vehicular network based on data coming from off-the-shelf IEEE 802.11p devices. The disadvantages of this experimental evaluation are the cost, the required time and the inability to reproduce -- since it is affected by physical parameters varying over time and it cannot be easily isolated and ignored.

A very complex real-world system is harder to model or might require increased resources to be simulated. Trial results can aid the design of a system by abstracting various parameters and introducing them as a priori knowledge in a simulation. For instance, authors in~\cite{gemv}, designed a simulator for geometry-based efficient propagation models for Vehicle-to-Vehicle (V2V) communications. Their simulator was based on an extensive trial campaign, in order to identify parameters such as the path loss exponent and the small-scale signal deviation for different distances and environments. Trial-based results though may vary between different devices, being related with the quality of the equipment used and the software accompanying it.

For the above reasons, it is obvious that various approximations, random and systematic errors are introduced during a system performance evaluation. A direct comparison between the simulated and real-world results will end up in a performance difference. To increase the accuracy of this scientific evaluation, in this paper we will establish a framework where real-world trials and simulations co-exist. Sharing knowledge between them, we will fine-tune a full-stack network simulator, enhancing the accuracy of our results and giving us the leverage of more precise experimentation later. For our trials, an open-source testbed will be used, consisting of single-board devices equipped with different wireless Network Interface Controllers (NICs) designed to be IEEE 802.11p compliant. The simulated results will be acquired using the VEINS network simulator~\cite{veins}. This is a vehicular networking framework based on Omnet++~\cite{omnetpp} and is compliant with the IEEE 802.11p/Wireless Access in Vehicular Environments (WAVE) standards.

The rest of the paper is organized as follows. In Sec.~\ref{sec:framework}, the hierarchical framework and the interrelationship between simulations and trials are described. The different entities of this framework, the relationship between them and the practical issues of the validation process are discussed. In Sec.~\ref{sec:procedure}, the procedure to fine-tune VEINS is analyzed, starting from an initial calibration isolating various parameters and moving towards a full-stack system optimization. Individually analyzing each scenario, their fine-tuned performance evaluation is examined in Sec.~\ref{sec:results}. Finally, Sec.~\ref{sec:conc} concludes the paper and provides future research avenues.

\section{Hierarchical Framework: Trials and Simulations}\label{sec:framework}
Consider an ITS consisting of a number of CAVs and Road Side Units (RSUs) on a road network. Different kinds of data are exchanged with respect to the safety- or infotainment-related applications and services running on the ITS.  For example, WAVE Short Messages (WSMs)~\cite{wave_standards}, are safety-critical messages encapsulating core information about CAVs (e.g. position, velocity, size, etc.). These messages are either broadcast every \SI{100}{\milli\second} or are triggered to announce road hazards. They are relatively short (\textasciitilde\SI{300}{}-\SI{800}{\byte}) and a high delivery rate and a low one-hop end-to-end delay are regarded as their Quality-of-Service (QoS) constraints.


Apart from WSMs, safety-critical applications will be key in future ITSs. For example, video-assisted overtaking or traffic monitoring applications are tested on CAVs~\cite{video}. These applications require the transmission of video streams encapsulated within UDP packets. Increased data rate and low jitter are their main QoS requirements, with a more forgiving bit error rate (BER) performance due to the adoption of Forward Error Correction (FEC) codes and the new generation higher efficiency video encoders.


Following the content-related QoS requirements, each application behaves differently under various physical environments. Three different environments can be found in vehicular networks (urban, suburban, rural)~\cite{rician_downs}. For example, urban environments are affected by blockages from the buildings, significantly attenuating the signal. Urban canyon behaviors can be introduced, under specific circumstances, waveguiding the signal. On the other hand, foliage is the main form of blockage in rural areas while vehicles tend to move faster introducing a higher Doppler Shift effect. Each scenario should be approached differently when simulated, adapting accordingly the various channel model characteristics.



\subsection{Co-operation and Co-existence of Trials and Simulations}

\begin{figure}[t]     
\centering
\includegraphics[width=0.9\columnwidth]{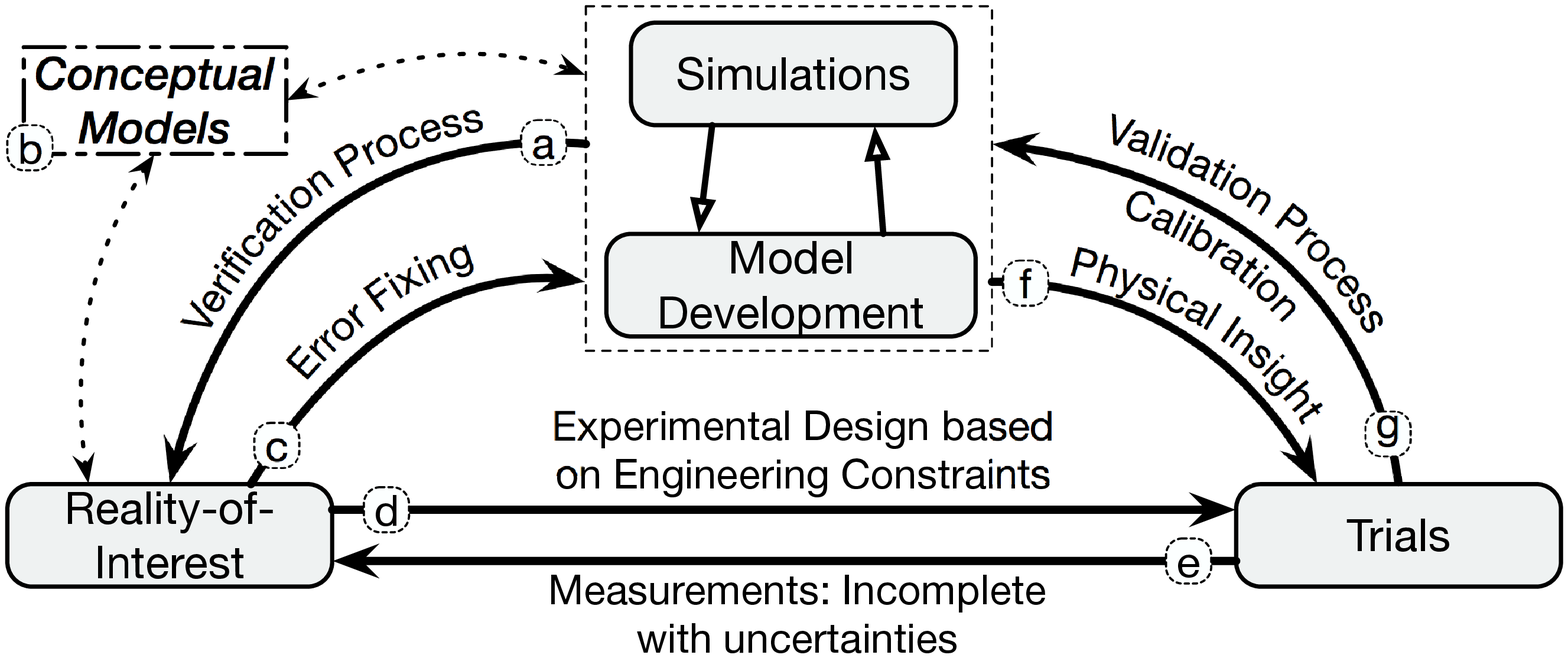}
    \caption{The interrelationship between the experiments, the simulations and the real world as well as the processes that connect them.}
    \label{fig:interrelationship}
\end{figure}

Cooperation between trials and simulations is mandatory to increase the accuracy of a system performance validation. Exchanging information between them can enhance the outcome, maximize the time utilization and minimize the cost. Establishing a framework between the \emph{reality-of-interest}, i.e. the part of the real world (e.g. a city, a neighborhood or a road) that we are interested in investigating, the \emph{trials} conducted and the \emph{simulation models} used, will help us better understand the requirements and the limitations of each one. The interrelationship between these three entities can be found in Fig.~\ref{fig:interrelationship}.

The assessment of the simulation accuracy can be divided in two phases, the \emph{validation} and the \emph{verification}~\cite{verify_validate}. The verification (Fig.~\ref{fig:interrelationship} -- a) is the confirmation that a model is correctly implemented and reflects the real world. This is confirmed by conceptual models -- abstractions of a system that characterize and standardize a network function, e.g. the OSI model (Fig.~\ref{fig:interrelationship} -- b). Verifying a model, the existing errors can be determined and fixed to assure that it matches specifications and assumptions with respect to the initial concept (Fig.~\ref{fig:interrelationship} -- c). For this work, we will not discuss the concept of the model verification for VEINS any further. For more, we kindly refer the reader to~\cite{veins_validation}.

A trial-based evaluation is limited by the engineering constraints (Fig.~\ref{fig:interrelationship} -- d). For example, a NIC might support decreased transmission power compared to the standard specifications, limiting the operational range of a device. What is more, trials suffer from uncertainties (e.g. the attenuation of the signal with respect to the weather is unpredictable, not easily measured and environment-dependent) affecting the reliability and the validity of the experiment. The replicability of the experiment is also a big concern. All roads are not the same and the devices have different specifications, so replicating an experiment is difficult. This leads to incomplete results (Fig.~\ref{fig:interrelationship} -- e) as it is impossible to validate all possible combinations. 

The validation of a theoretical model assesses the fidelity that a model reproduces the state and behavior of the real world (Fig.~\ref{fig:interrelationship} -- f). Verification usually precedes the validation of a model. The models can be validated with simple experiments isolating the external factors affecting the performance. To achieve a meaningful representation of the real world, a simulator should be fine-tuned at first (Fig.~\ref{fig:interrelationship} -- g) using inputs from measured results (e.g. path loss exponent) or applying weights at the output to minimize the divergence error.  
A direct comparison between the absolute values of the experimental results should be avoided. The validation process should focus on the trends of performance between the different experiments (e.g. both the simulated and trial-based results have a relative degradation when one parameter is changed).

\subsection{Hierarchical Validation of a Simulation Model}\label{sub:practical}
Differences between trial and simulated results can arise from various reasons. Typical examples are \emph{measurement errors} (e.g. calibration errors, noise or data acquisition methods, etc.), \emph{formulation errors} (e.g. incorrect channel models) or \emph{numerical errors} (e.g. overflow of integers, subtraction of floating points, etc.). Two different error categories exist. Firstly, the \emph{random} errors, affecting the relative precision of a model or a measurement. Secondly, the \emph{systematic} errors affecting the absolute value of a result, being repeatable though and therefore, easily predicted.

In order to achieve the required level of accuracy, the validation procedure should be carried out throughout the entire development process following a hierarchical approach~\cite{validation}. Fragmenting the problem into smaller entities and solving them individually, the necessary level of precision can be achieved without increasing the complexity. Generally, the trials should test crucial features of the simulation models, such as the impact of the considered assumptions or the simplifications. On the other hand, simulations allow incremental validation towards a ``real-world-like'' system.



\section{Fine-Tuning VEINS Network Simulator}\label{sec:procedure}
Designing a real-world ITS solution, all the above should be taken into account as well as the initial conditions, the boundaries and the trends of the performance. Isolating the characteristics that disruptively affect the performance we can isolate the systematic and the random errors approaching an "ideal-like" system. Using a network simulator, we can further validate different scenarios based on our initial configuration. To do so, the fine-tuning process of VEINS requires a detailed study on the available hardware, the performance metrics used, an insight into how the simulator operates and what are the differences of the real-world. This study will be discussed in the next subsections. 

\begin{figure}[t]     
\centering
\includegraphics[width=0.85\columnwidth]{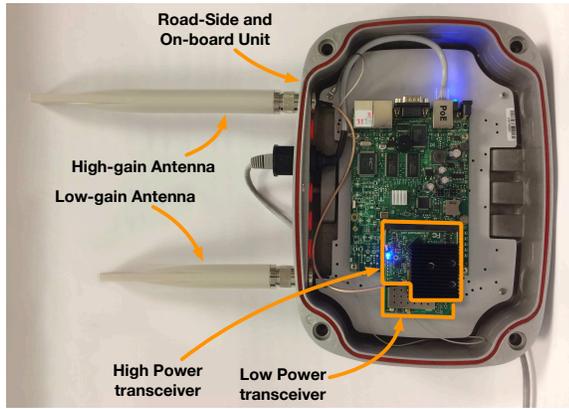}
    \caption{Low-latency Linux Kernel implementation of IEEE 802.11p/DSRC units. The figure shows units designed for RSU and OBU deployment.} 
    \label{fig:units}
\end{figure}

\subsection{Description of the Experimental Setup}\label{sub:devices}
For our experimental validation, we prototyped an open-source IEEE 802.11p/DSRC testbed (Fig.~\ref{fig:units}) meeting the following requirements:
\begin{itemize}
    \item \emph{Easily customisable}.
    \item \emph{Low cost} with the potential to be massively deploy later.
    \item \emph{Open-source operating system} providing enough flexibility for future developments.
    \item \emph{Dual-operation as RSUs and On-Board Units (OBUs)}.
    \item \emph{Weatherproof}.
\end{itemize}

To meet these requirements, each device was equipped with a Mikrotik RB433 single-board computer (CPU \SI{300}{\mega\hertz}, \SI{64}{\mega\byte} RAM, \SI{64}{\mega\byte} storage space, x3 Ethernets, x3 MiniPCI slots)~\cite{rb433}. Two IEEE 802.11a NICs were used for redundancy in the wireless links. The first one was a Mikrotik R52H~\cite{r52h}, regarded in this work as a low-power (LP) NIC, with transmission power of up to \SI{25}{\dBm} and connected to a dipole antenna of \SI{7}{\dBi} gain at \SI{5.9}{\giga\hertz}. The second model was a Mikrotik R5SHPn~\cite{R5SHPn}, operating as the high-power (HP) transceiver in our experiments (\SI{29}{\dBm} maximum transmission power). This NIC was connected to a \SI{9}{\dBi}-gain dipole antenna.

\begin{figure}[t]     
\centering
\includegraphics[width=0.75\columnwidth]{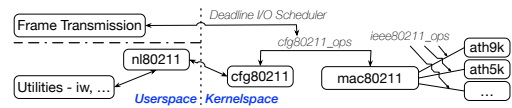}
    \caption{Linux Kernel Modules that have been modified to implement a IEEE 802.11p/DSRC stack.}
    \label{fig:drivers}
\end{figure}

A low-latency OpenWRT Linux distribution was used\footnote{OpenWRT Barrier Breaker Release no. 14.07 - https://openwrt.org/} as the operating system. The different Atheros chipsets of each transceiver (AR5414 for the LP and AR9220 for the HP) required the use of two different  Atheros drivers (ath5k for the LP and ath9k for the HP). Both were modified accordingly to enable IEEE 802.11p compatibility. The Linux kernel modules that we modified have been summarized in Fig.~\ref{fig:drivers}. The software modules \emph{cfg80211} and \emph{nl80211} act as interfaces between the user and kernel space, \emph{mac80211} is the general driver framework, and \emph{iw} is the NIC configuration utility. Furthermore, \emph{cfg80211\_ops} and \emph{ieee80211\_ops} define the operations and the callbacks between the different blocks. The Outside the Context of a BSS (OCB) mode was enabled in the MAC layer, allowing the NICs to operate without being associated with an access point and the \emph{iw} utility was modified accordingly to include the new commands for using OCB mode. Finally, the \SI{5.9}{\giga\hertz} band was added in the regulatory domain.

Each device can be connected with a GPS dongle via its USB interface. Finally, a beaconing interface was developed that was able to acquire viable vehicle information (e.g. length, position, emissions, etc.) from an Engine Control Unit (ECU) and broadcast them to surrounding devices.



\subsection{Initial V2X Calibration Scenario}\label{sub:initial}

Consider a scenario with an ideal channel (no reflections, free-space path loss) between two ideal stationary vehicles (isotropic antennas, zero hardware attenuation). As this ideal system does not exist in the real world, the experimental setup was initially demonstrated inside an anechoic chamber (\SI{8.5}{\meter}$\times$\SI{4.5}{\meter}$\times$\SI{5}{\meter}) using both the HP and LP transceivers. A number of experiments was conducted at a distance of \textasciitilde\SI{6.5}{\meter} for each transceiver and each MCS using two devices, one acting as a RSU and the other as an OBU. 
A UDP data stream, transmitted from the OBU to the RSU and generated using iPerf traffic generator\footnote{iPerf Traffic Generator - https://iperf.fr}, and a periodic beacon every \SI{100}{\milli\second}, were used to saturate the channel. As known, the network level performance is affected by the signal-to-interference-plus-noise ratio (SINR) and the sensitivity levels for each MCS. The SINR degrades based on the disruptive characteristics of the channel (e.g. distance attenuation, multipath, antenna misalignment, etc.) and the devices (e.g. thermal noise, etc.). Using an anechoic chamber, we conducted the experiments under near-optimal conditions (SINR greater than the sensitivity level) and therefore the optimal performance was achieved.


The same scenario was designed in VEINS as well. By partitioning the design process into smaller problems, we managed to achieve the required level of similarity. Firstly, we considered the IEEE 802.11p Physical Layer (PHY) frame, which consists of three fields~\cite{physical_layer}: (i) The \emph{Preamble} marks the beginning of the PHY frame, is responsible for the appropriate antenna selection and corrects the timing and frequency offsets, (ii) the \emph{Signal} field (SIG) specifies the frame rate and length and (iii) The \emph{Data} field consisting of the Physical layer Service Data Unit (PSDU) that encapsulates the MAC frame, the Physical Layer Convergence Procedure (PLCP) Service, and a Tail field. The Data field can also be padded with extra bits so its length is a multiple of the coded bits in an OFDM symbol.
The above are transmitted using BPSK $\nicefrac{1}{2}$ Modulation and Coding Scheme (MCS). 
The length of each field can be found in Tab.~\ref{tab:parameters}.

\begin{figure*}[t]     
\centering
\includegraphics[width=0.92\textwidth]{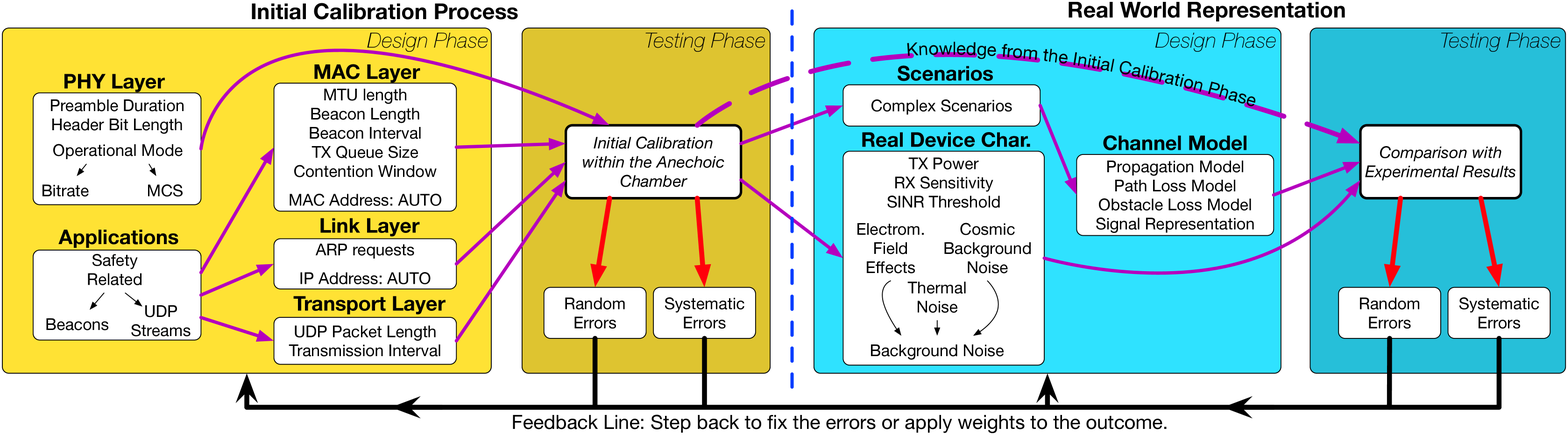}
    \caption{Hierarchical Validation Process. Each entity is individually fine-tuned to achieve high-fidelity results after the calibration of the entire system. The fine-tuning is a two-phase process starting from an initial calibration and moving to a more complex real-world representation.} 
    \label{fig:hierarchicalmodel}
\end{figure*}

In VEINS, the duration of the Preamble and SIG is controlled by the parameter \emph{preambleDuration}, while the bit length of the Data field is set equal to \emph{headerBitLength} (see Fig.~\ref{fig:hierarchicalmodel}). The simulated PHY bitrates are governed by the wireless interface operational mode -- namely, \emph{opMode} that has been set to ``\emph{p}'', in this case.
Each MCS should be manually configured for each individual simulation and matched with the appropriate PHY bitrate using pairs of the simulation parameters \emph{bitrate} and \emph{modulation}. Other PHY parameters that should be set within VEINS are the channel \emph{bandwidth}, the \emph{carrierFrequency}, the \emph{antennaType} (\emph{ConstantAntennaGain} in this case) and the \emph{gain}.

The multi-channel operation introduced in the WAVE 1609.4 standard~\cite{wave_standards} was not considered in order to identify the maximum performance under saturation conditions. Furthermore, the RTS threshold -- namely \emph{rtsThreshold}, was set to a value greater than the frame size. This ensured that the RTS/CTS procedure was disabled. The MAC layer backoff times are drawn from a Contention Window (CW) starting from $CW_{\mathrm{min}}$ (\emph{cwMinData} and \emph{cwMaxData}). The values chosen for our setup (Tab.~\ref{tab:parameters}) where proven to be optimal for vehicular communications~\cite{contention}. The length of the MAC TX queue size is capped by the driver (in the ath5k case), so the same value was considered in the simulation as well (\emph{maxQueueSize} parameter within VEINS).

The VEINS parameter \emph{sentInterval} sets the interval between the generation of two consecutive UDP packets. In order to saturate the channel, a very precise interval was chosen for each MCS. Suitable values were found with a trial and error method to fully utilize the channel without having packets discarded from the MAC TX queue. All the simulation and experimental parameters can be found in Tab.~\ref{tab:parameters} and Sec.~\ref{sub:devices}. 

\begin{table}[t]
\renewcommand{\arraystretch}{1.07}
\centering
    \caption{Simulation and Experimental Parameters.}
    \begin{tabular}{r|rl}

    \textbf{Parameter}           & \textbf{Value}    & \\ \hline \hline
    Experiment/Simulation Time   & \SI{10}           & \SI{}{\second}  \\
    Carrier Frequency            & \SI{5.9}          & \SI{}{\giga\hertz}  \\
    Bandwidth            & \SI{10}          & \SI{}{\mega\hertz}  \\
    MTU                          & \SI{1500}         & \SI{}{\kilo\byte} \\  
    UDP Packet Length            & \SI{8192}         & \SI{}{\kilo\byte}  \\
    Beacon Length                & \SI{500}          & \SI{}{\byte}  \\ 
    Beacon Interval              & \SI{100}          & \SI{}{\milli\second}  \\
    Preamble Duration            & \SI{32}           & \SI{}{\micro\second}  \\ 
    SIG Duration                 & \SI{8}            & \SI{}{\micro\second}  \\ 
    PLCP Service Length          & \SI{16}           & \SI{}{\bit}  \\ 
    Tail Length (Data Field)     & \SI{6}            & \SI{}{\bit}  \\ 
    $CW_{\mathrm{min}}, CW_{\mathrm{max}}$ & $\left[15, 1023\right]$ & \\
    TX MAC queue size            & 50                & \\
    Background Noise             & $\mathcal{N}\left( -110, 3 \right)$ & \SI{}{\dBm} \\
    Connector and Cable Losses   & 3                 & \SI{}{\dBm}
    \end{tabular}
\label{tab:parameters}
\end{table}

\begin{figure*}[t]     
\centering
\includegraphics[width=0.88\textwidth]{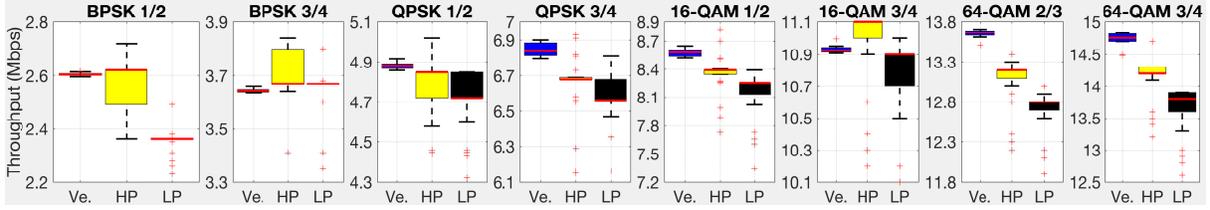}
    \caption{Values of throughput obtained from VEINS (initial calibration), the HP and LP transceivers, for different MCSs.}
    \label{fig:initialThroughput}
\end{figure*}
\begin{figure*}[t]     
\centering
\includegraphics[width=0.88\textwidth]{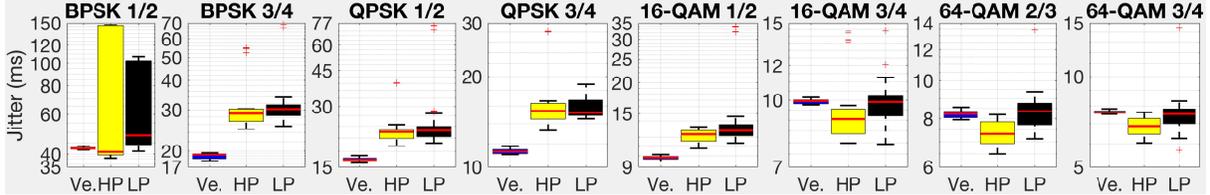}
    \caption{
    Values of jitter associated to the UDP stream. Obtained from VEINS (initial calibration), the HP and LP transceivers for different MCSs.}
    \label{fig:initialJitter}
\end{figure*}

The results for the above calibration scenario are shown in Figs.~\ref{fig:initialThroughput} and~\ref{fig:initialJitter}. The central rectangle is the interquartile region (IQR) between the first and the third quartile, while the line within represents the median. The whiskers constitute the maximum and the minimum values and the asterisks show the outliers. A value is regarded as an outlier if it outside $\pm 2.7\sigma$ (99.3\% percentage coverage of the normally distributed samples).


Fig.~\ref{fig:initialThroughput} shows the reception throughput measured at the transport layer. As mentioned in Sec.~\ref{sub:practical}, a meaningful comparison should focus on the trends. 
Therefore, a trend can be seen in the performance whereby the simulation results are slightly better for some MCSs (e.g. QPSK \nicefrac{3}{4}) while for others (e.g. BPSK \nicefrac{1}{2}) they are almost identical. The deviation in the mediam is of the order of up to \textasciitilde\SI{0.5}{\Mbps} for the HP device and \textasciitilde\SI{1}{\Mbps} for the LP transceiver. Overall, we observed that the LP transceiver has a worse throughput performance (median values decreased by \textasciitilde5\%) compared to the HP one; following the same trend for all MCSs. This difference is due to the operation of the different drivers used.  

Fig.~\ref{fig:initialJitter} compares the inter-arrival jitter performance. The jitter, as defined in RFC 1889~\cite{rtp}, is the statistical variance of the inter-arrival time between packets. 
Comparing the absolute values of the results, we see a huge difference between the trials and the simulations. However, comparing the relative variation in the jitter performance for the different MCSs, it is shown that in both cases, the results follow a similar trend starting with an increased jitter for the lower MCSs, and having a better performance as the bitrate is increased.
To that extent, the jitter values measured with VEINS have been multiplied by $56000$ in order to get the same order of magnitude with the ones obtained by the LP and HP transceivers. The huge difference in the absolute values was somehow expected as our devices are built upon a single-core CPU, which executes tasks with the same priority according to the Linux \emph{Deadline I/O Scheduler} -- thus the CPU cannot fetch/push data streams towards the transceivers at a constant I/O rate. Since in VEINS this issue is not present, the simulated and measured jitter performance may vary significantly.

With regards to the framework introduced in Sec.~\ref{sec:framework}, we fragmented our system into smaller problems, fine-tuning individually each of them and managed to build an ideal subsystem able to achieve similar performance with minor deviations between the trial and the simulated results. In the next sections, we will introduce more complex entities such as the real device profiles and the channel behavior approaching a better representation of the real world. 



\subsection{Moving towards a Realistic Representation of the World}
The surrounding environment plays a key role in the performance of vehicular networks. Each environment has its own characteristics, therefore a theoretical analysis is required for the signal degradation due to the different channels. The channel behavior within VEINS lies beneath the \emph{physical medium model}. This model is further split into different submodules (Fig.~\ref{fig:hierarchicalmodel}). 

At first, the propagation model (\emph{propagationType}) describes the propagation time within the channel. Considering \emph{constantSpeed} as our propagation model, the propagation time is proportional to the distance travelled. To accurately represent the signal and its fluctuation, a \emph{DimensionalAnalogModel} was utilized, meaning that the signal power deviation is represented over both time and frequency.

The long-term signal degradation in the real-world depends on the distance, the carrier frequency, the device positioning, etc. In VEINS, the simulation parameter \emph{pathLossType} describes the path loss model that is responsible for computing the power reduction based on the traveled distance, the velocity factor, the carrier frequency and the path loss exponent for each environment. The short-term signal degradation, affected by the multipath distortion caused by the surrounding buildings can be described within VEINS using small-scale fading models (e.g. Nakagami, Rician, etc.) and fine-tuning their individual parameters (shape-factor for Nakagami, K-factor for Rician, etc.) accordingly.
An obstacle loss model can be added to an existing path loss model. If so, the parameter \emph{ObstacleLossType} specifies the material absorption when a ray is traveling within an object. The physical obstacles can be listed using an XML file processed by the \emph{physical environment model} within VEINS.

\subsection{Integration of Real Device Profiles in VEINS}
After configuring the parameters related to the signal propagation, it is key to consider the different characteristics of each simulated NIC. Unfortunately, commercial-off-the-shelf (COTS) devices have frequently to be treated as a ``black box'' as many of their physical level performance characteristics are unknown and not easily measurable. Therefore, we will base some of our simulation parameters on speculations based on the datasheet of each considered transceiver.

At first, the SINR can fluctuate from random effects such as the thermal noise, the cosmic background noise, electromagnetic field effects, etc. These effects are not predictable and do not particularly come from a specific source to be isolated. VEINS represents this signal variation with a background noise model (\emph{backgroundNoiseType}), configured using the \emph{backgroundNoise} parameter following a Normal distribution. Also, the cables and the connectors in a system introduce a systematic attenuation described as the \emph{systemLoss} that was measured within our laboratory. 

According to the manufacturer datasheet, we have access to the transmission power given at \SI{20}{\mega\hertz} of channel bandwidth, for each MCS. However, the IEEE 802.11p bandwidth is equal to \SI{10}{\mega\hertz}. From the energy-per-symbol-to-noise power spectral density equation it follows that:
\begin{equation}
\frac{E_s}{N_0} = \frac{C}{N} \, \frac{B}{f_s}
\end{equation}
where $E_s$ is the energy per symbol, $N_0$ is the noise power, $C/N$ is the carrier-to-noise ratio, $B$ is the channel bandwidth and finally $f_s$ is the symbol rate. In our case, the only non-constant variable is $B$. Therefore, for a \SI{10}{\mega\hertz} channel, the $E_s/N_0$ ratio is expected to be twice as much as that measured using a \SI{20}{\mega\hertz} channel.
As $N_0$ is measured per unit of bandwidth (per \SI{}{\mega\hertz}), it follows that $E_s$ is doubled. 
Finally, knowing the number of bits per symbol of each MCS, we can infer the maximum transmission power for a \SI{10}{\mega\hertz} channel. These values are summarized in Tab.~\ref{tab:assumptions}.

With regards to the sensitivity of the receiver, from the Minimum Operational Sensitivity (MOS) relation, we know that:
\begin{equation}\label{eq:MOS}
\mathrm{MOS} = \frac{SINR_{\mathrm{thr}} \, k \, T_{\alpha} \, B \, \left(NF \right)}{G_{\mathrm{rx}}}
\end{equation}
where $SINR_{\mathrm{thr}}$ is the minimum SINR needed to process (not just detect) a signal, $NF$ is the noise figure, $k$ is Boltzmann's Constant, $T_{\alpha}$ is the effective noise temperature referred at the input of the receiver, and $G_{\mathrm{rx}}$ is the isotropic antenna gain. Obviously, $SINR_{\mathrm{thr}}$ depends not just on the NIC but also on the MCS in use. 
Authors in~\cite{receiver_sens} measured the $SINR_{\mathrm{thr}}$ under a V2I scenario for two different antenna heights. Their lower antenna configuration was very similar to ours, so their $SINR_{\mathrm{thr}}$ results will be utilized for our scenarios. 
Knowing the $SINR_{\mathrm{thr}}$, the only variable in~\eqref{eq:MOS} is $B$. Therefore, halving $B$, the MOS will be doubled. Finally, the antenna gain values were taken directly from the manufacturer datasheet. All the values are presented in Tab.~\ref{tab:assumptions}. 

\begin{table}[t] 
\renewcommand{\arraystretch}{1.07}
\centering
    \caption{Simulation Parameters based on the Manufacturer Datasheet.}
    \begin{tabular}{r||cc|cc|cc|m{0.4cm}}

    \multirow{2}{*}{\textbf{Modulat.}} & \multicolumn{2}{c|}{\textbf{TX power}} & \multicolumn{2}{c|}{\textbf{RX sensitivity}} & \multicolumn{2}{c}{\boldmath$SINR_{\mathrm{thr}}$}\cite{receiver_sens} & \multirow{2}{*}{\textbf{Units}} \\ 
                 & \textbf{LP} & \textbf{HP}   & \textbf{LP}     & \textbf{HP}     &  \scriptsize\textbf{\nicefrac{1}{2} MCS} & \scriptsize\textbf{\nicefrac{3}{4} MCS} &   \\ \hline\hline
    BPSK         & \SI{27}     & \SI{28} & \SI{-93}  & \SI{-93}  & \SI{10}       & \SI{15} & \SI{}{dBm} \\
    QPSK         & \SI{26}     & \SI{27} & \SI{-85}  & \SI{-88}  & \SI{10}       & \SI{15} & \SI{}{dBm} \\    
    16-QAM       & \SI{25}     & \SI{26} & \SI{-80}  & \SI{-84}  & \SI{17}       & \SI{17} & \SI{}{dBm} \\     
    64-QAM       & \SI{24}     & \SI{24} & \SI{-73}  & \SI{-80}  & \SI{20}       & \SI{25} & \SI{}{dBm} \\        
    \end{tabular}
\label{tab:assumptions}
\end{table}

\section{Performance Evaluation and Macroscopic View}\label{sec:results}

\subsection{Scenarios and channel analysis}

In Sec.~\ref{sec:procedure}, the full-stack calibration process for VEINS was discussed. As mentioned, the different environments significantly change the behavior of the system. To that extent, three different scenarios were designed and evaluated. As in the initial calibration scenario, we considered one RSU and one OBU devices, stationary during the experiments having their performance being evaluated for different distances, MCSs and for both the HP and LP transceivers. Again a UDP stream and a periodic beaconing transmitted from the OBU to the RSU saturate the channel to evaluate the network throughput (as described in Sec.~\ref{sec:framework} and~\ref{sub:initial}). The two different transceiver configurations were simulated within VEINS using the parameters from Tabs.~\ref{tab:parameters} and~\ref{tab:assumptions} based on the analysis preceded in the previous sections.

\begin{figure}[t]     
\centering
\includegraphics[width=1\columnwidth]{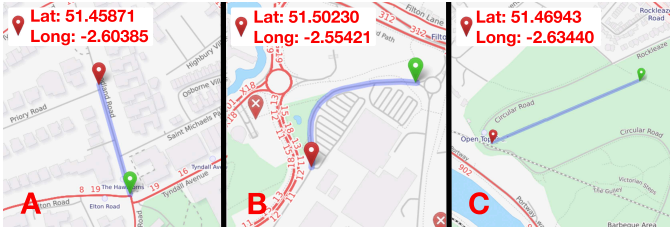}
    \caption{The three different scenarios were conducted around the city of Bristol, UK. a) the urban, b) the suburban, c) the rural scenario.} 
    \label{fig:scenarios}
\end{figure}

The first scenario (see Fig.~\ref{fig:scenarios} -- A) is an urban road with buildings on both sides. The devices are positioned on the pavement and there is always a Line-of-Sight (LOS). The buildings surrounding the devices cause multipath distortion. However, since the devices are always in LOS and the experiment was conducted at a relatively short distance, a Rician fading model was considered with a $K$-factor $k=\SI{3.36}{\dB}$ and a path loss exponent $\alpha = 2.3$~\cite{rician}.

The second scenario (see Fig.~\ref{fig:scenarios} -- B) is a suburban area on a bent and sloppy road with foliage in between the devices and a few buildings on one side of the road. LOS existed between the devices for the first \textasciitilde\SI{50}{\meter}. For the rest of the experiment, the RSU was hidden behind the road slope and the vegetation. This scenario can be split into two different subscenarios. Up to \SI{50}{\meter}, we considered a Rician fading model with  $k=\SI{2.45}{\dB}$ and $\alpha = 2.3$~\cite{rician}. For the NLOS part, we refer to Rayleigh fading model with $\alpha = 2.5$.

The third scenario refers to a rural environment (see Fig.~\ref{fig:scenarios} -- C). Both the RSU and OBU are always in LOS, and no high buildings or other objects were surrounding the devices apart from some foliage. Therefore, the impact of multipath was minimum. As such, a Rician fading model was considered, with $k=\SI{8}{\dB}$ and $\alpha = 2.2$~\cite{rician_downs}. For this scenario, only the HP transceiver was used. 

\subsection{Performance Evaluation}
During the experiments carried in scenarios A, B and C, the RSU and OBU were fitted on a tripod at \textasciitilde\SI{1.8}{\meter} height.
A consistent setup has been simulated in VEINS. During both the trials and simulations, each performance metric we measured is the result of an average of multiple experiments.
The ARP probe was disabled by manually inserting the addresses of the devices in their respective ARP tables. Due to limited space, the most meaningful results for each distance and MCS will be shown while the rest will be described within the text. 

\begin{figure*}[t]     
\centering
\includegraphics[width=0.94\textwidth]{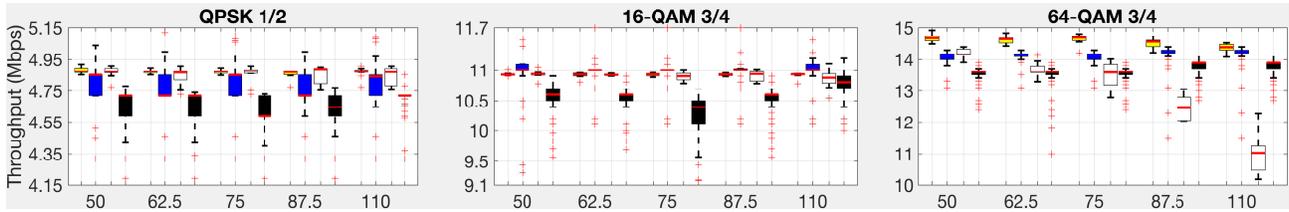}
    \caption{Values of throughput as a function of the distance between RSU and OBU. Results refer to an urban scenario. Each quartet represents the results for a single position with the order (from left to right): 1) VEINS-HP, 2) Trials-HP, 3) VEINS-LP, 4) Trials-LP. 
    }
    \label{fig:woodland}\vspace{-2mm}
\end{figure*}

With regards to the urban scenario, Fig.~\ref{fig:woodland} shows the communication throughput that can be sustained by the OBU as a function of distance between RSU and OBU, for each MCSs.
For lower modulations (BPSK, QPSK) and all distances, the same trend and performance were seen as in the calibration process (Fig.~\ref{fig:initialThroughput}). 
Again, the difference in the performance between the transceivers, not observed within VEINS, is due to the different drivers used. 
Increasing the MCS and the distance separating the devices, even though the median values remain similar to what shown in Fig.~\ref{fig:initialThroughput}, the introduced multipath distortion leads to a larger number of outliers.
For 16-QAM \nicefrac{1}{2} and \nicefrac{3}{4}, it was seen that as the distance is increased, the SINR drop starts being observed within VEINS for the LP configuration (e.g. \SI{110}{\meter}) having a different behavior compared to the trials. Finally, for 64-QAM \nicefrac{2}{3} and \nicefrac{3}{4}, the signal received from the LP transceiver within VEINS is significantly attenuated. The performance degradation is about \SI{1}{\Mbps} compared to the calibration scenario (\SI{50}{\meter}) reaching up to \SI{3}{\Mbps} at \SI{110}{\meter}. 

\begin{figure*}[t]     
\centering
\includegraphics[width=0.94\textwidth]{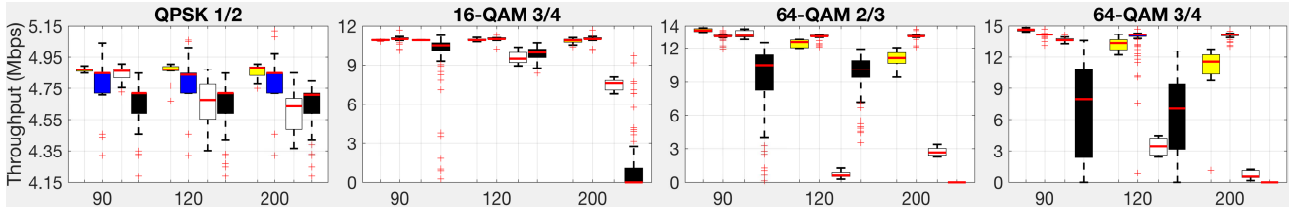}
    \caption{Values of throughput as a function of the distance between RSU and OBU. Results refer to a suburban scenario. Each quartet follows the same order as in Fig.~\ref{fig:woodland}}.
    \label{fig:uwe}\vspace{-2mm}
\end{figure*}

Fig.~\ref{fig:uwe} shows the communication throughput measured in the case of the suburban scenario, for different MCSs.
For distances of \SI{30}{\meter} and \SI{60}{\meter}, we observe that the results follow the same trend as in the urban case (see Fig.~\ref{fig:woodland}). In particular, despite the RSU being hidden after \SI{50}{\meter} due to the road slope, the overall communication throughput was not severely impacted. For greater distances and the HP scenario, VEINS behaves slightly worse compared to the actual device. However, for the LP scenario, the actual device achieved less throughput compared to the simulated result. Especially for 64-QAM \nicefrac{2}{3} and \nicefrac{3}{4} and a distance of \SI{200}{\meter}, our LP transceiver achieved zero throughput during the trials whereas the VEINS result is around \SI{3}{\Mbps} and \SI{1.2}{\Mbps} respectively. This is due to the BER calculation within VEINS that is approximated based on a Gaussian error function not exactly reflecting the reality. 

\begin{figure*}[t]     
\centering
\includegraphics[width=0.94\textwidth]{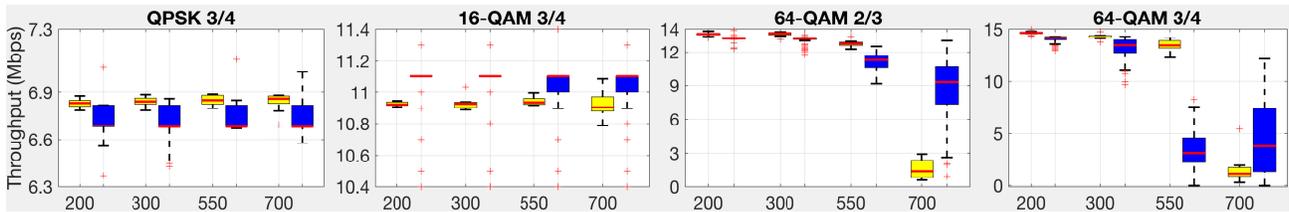}
    \caption{Values of throughput as a function of the distance between RSU and OBU. Results refer to a rural scenario. Each boxplot pair is: 1) VEINS-HP, 2) Trials-HP.
    }
    \label{fig:downs}\vspace{-5mm}
\end{figure*}

Fig.~\ref{fig:downs} refers to the rural scenario.
Again, for lower modulation schemes (BPSK, QPSK), the same trend was observed as before. For higher MCS, VEINS again exhibits a sharp performance degradation when the distance increases. This is clearer at 64-QAM \nicefrac{2}{3} and \nicefrac{3}{4} where VEINS outperforms the trial performance at \SI{550}{\meter} as expected from the trend observed. However, it is significantly worse at \SI{700}{\meter}. 

\section{Conclusions, Lessons Learned and Future Work}\label{sec:conc} 

In this paper, we described the process to follow in order to calibrate a full-stack network simulator for vehicular communications. In our study, we considered VEINS as the network simulator of interest. At first, we introduced an initial calibration process where an ``ideal-like'' scenario was evaluated. Two different transceivers were used to demonstrate the difference emanating from dissimilar hardware and how it can impact the experimental results. Specifically, we focused on result trends and identified the reasons that they exist.

To that extent, a trend in the performance was shown for both the real and the simulated scenario. For some MCSs (e.g. BPSK \nicefrac{1}{2}, QPSK \nicefrac{1}{2}, etc.) the network throughput is almost identical whereas for the others there is a slight deviation of up to \textasciitilde\SI{1}{\Mbps} (e.g. 64-QAM \nicefrac{3}{4}, etc.). The different drivers and hardware introduced a dissimilarity between the HP and the LP transceivers, something not observed within VEINS as the differences in the hardware cannot be easily simulated. A similar trend was seen in the jitter performance evaluation as well. The huge difference between the absolute values due to the Linux scheduling algorithm serves as a proof of capacity that the direct comparison between simulated and actual trial results should be avoided.

After the initial calibration, VEINS was further fine-tuned to achieve high-fidelity results for more complex scenarios. Three different vehicular environment scenarios were evaluated with respect to the network throughput. The trends identified, showed that for the HP transceiver VEINS has a sharper performance degradation when the distance is increased whereas, for the LP one, it behaves smoother. As discussed, some of the simulation parameters are based on various assumptions and therefore this difference in the performance is introduced. Finally, we concluded that the proposed tuning process allows VEINS to deliver high-fidelity simulation results of IEEE 802.11p/DSRC links, which will be pivotal in city-scale simulation scenarios -- scenarios impossible to be calibrated otherwise.

\bibliographystyle{IEEEtran}
\bibliography{bib.bib,IEEEabrv}

\begin{thebibliography}{10}
\providecommand{\url}[1]{#1}
\csname url@samestyle\endcsname
\providecommand{\newblock}{\relax}
\providecommand{\bibinfo}[2]{#2}
\providecommand{\BIBentrySTDinterwordspacing}{\spaceskip=0pt\relax}
\providecommand{\BIBentryALTinterwordstretchfactor}{4}
\providecommand{\BIBentryALTinterwordspacing}{\spaceskip=\fontdimen2\font plus
\BIBentryALTinterwordstretchfactor\fontdimen3\font minus
  \fontdimen4\font\relax}
\providecommand{\BIBforeignlanguage}[2]{{%
\expandafter\ifx\csname l@#1\endcsname\relax
\typeout{** WARNING: IEEEtran.bst: No hyphenation pattern has been}%
\typeout{** loaded for the language `#1'. Using the pattern for}%
\typeout{** the default language instead.}%
\else
\language=\csname l@#1\endcsname
\fi
#2}}
\providecommand{\BIBdecl}{\relax}
\BIBdecl

\bibitem{fully_autonomy}
J.~Levinson \emph{et~al.}, ``Towards {Fully} {Autonomous} {Driving}: {Systems}
  and {Algorithms},'' in \emph{IEEE Intel. Veh. Symp. IV}, Jun. 2011, pp.
  163--168.

\bibitem{broadband}
P.~Demestichas \emph{et~al.}, ``Intelligent {5G} {Networks}: {Managing} {5G}
  {Wireless}/{Mobile} {Broadband},'' \emph{IEEE Veh. Technol. Mag.}, vol.~10,
  no.~3, pp. 41--50, Sep. 2015.

\bibitem{TassiTVT}
A.~Tassi \emph{et~al.}, ``{Modeling and Design of Millimeter-Wave Networks for
  Highway Vehicular Communication},'' \emph{{IEEE} Trans. Veh. Technol.}, Aug.
  2017.

\bibitem{dense_network}
C.~Han, M.~Dianati, and M.~Nekovee, ``Effective {Decentralised}
  {Segmentation}-based {Scheme} for {Broadcast} in {Large}-scale {Dense}
  {VANETs},'' in \emph{IEEE WCNC 2016}, Apr. 2016, pp. 1--6.

\bibitem{adaptive_beaconing}
C.~Sommer, O.~K. Tonguz, and F.~Dressler, ``Adaptive {Beaconing} for
  {Delay}-{Sensitive} and {Congestion}-aware {Traffic} {Information}
  {Systems},'' in \emph{Proc. IEEE VNC 2015}, Dec. 2010, pp. 1--8.

\bibitem{experiments}
F.~A. Teixeira \emph{et~al.}, ``Vehicular {Networks} {Using} the {IEEE} 802.11p
  {Standard}: {An} {Experimental} {Analysis},'' \emph{Vehicular
  Communications}, vol.~1, no.~2, pp. 91 -- 96, 2014.

\bibitem{gemv}
M.~Boban, J.~Barros, and O.~Tonguz, ``Geometry-{Based} {Vehicle}-to-{Vehicle}
  {Channel} {Modeling} for {Large}-{Scale} {Simulation},'' \emph{IEEE Trans.
  Vehic. Techn.}, vol.~63, no.~9, pp. 4146--4164, Nov. 2014.

\bibitem{veins}
C.~Sommer, R.~German, and F.~Dressler, ``Bidirectionally {Coupled} {Network}
  and {Road} {Traffic} {Simulation} for {Improved} {IVC} {Analysis},''
  \emph{IEEE Trans. Mobile Comput.}, vol.~10, no.~1, pp. 3--15, Jan. 2011.

\bibitem{omnetpp}
\BIBentryALTinterwordspacing
``Omnet++ {Discrete} {Event} {Simulator}.'' [Online]. Available:
  \url{https://omnetpp.org}
\BIBentrySTDinterwordspacing

\bibitem{wave_standards}
S.~A.~M. Ahmed, S.~H.~S. Ariffin, and N.~Fisal, ``Overview of {Wireless}
  {Access} in {Vehicular} {Environment} ({WAVE}) {Protocols} and {Standards},''
  \emph{Indian Journal of Science and Technology}, vol.~6, no.~7, 2013.

\bibitem{video}
E.~Belyaev \emph{et~al.}, ``The {Use} of {Automotive} {Radars} in
  {Video}-{Based} {Overtaking} {Assistance} {Applications},'' \emph{IEEE Trans.
  Intell. Transp. Syst}, vol.~14, no.~3, pp. 1035--1042, Sep. 2013.

\bibitem{rician_downs}
S.~Zhu \emph{et~al.}, ``Probability {Distribution} of {Rician} {K}-{Factor} in
  {Urban}, {Suburban} and {Rural} {Areas} {Using} {Real}-{World} {Captured}
  {Data},'' \emph{IEEE Transactions on Antennas and Propagation}, vol.~62,
  no.~7, pp. 3835--3839, Jul. 2014.

\bibitem{verify_validate}
R.~G. Sargent, ``An {Introductory} {Tutorial} on {Verification} and
  {Validation} of {Simulation} {Models},'' in \emph{2015 Winter Simulation
  Conference (WSC)}, Dec. 2015, pp. 1729--1740.

\bibitem{veins_validation}
F.~Klingler, F.~Dressler, and C.~Sommer, ``Ieee 802.11p {Unicast} {Considered}
  {Harmful},'' in \emph{IEEE VNC 2015}, Dec. 2015, pp. 76--83.

\bibitem{validation}
P.~J. Roache, \emph{Verification and {Validation} in {Computational} {Science}
  and {Engineering}}.\hskip 1em plus 0.5em minus 0.4em\relax Hermosa
  Albuquerque, NM, 1998, vol. 895.

\bibitem{rb433}
\BIBentryALTinterwordspacing
``Mikrotik {RB433} {Data} {Sheet}.'' [Online]. Available:
  \url{https://mikrotik.com/product/RB433}
\BIBentrySTDinterwordspacing

\bibitem{r52h}
\BIBentryALTinterwordspacing
``Mikrotik {R52H} {Data} {Sheet}.'' [Online]. Available:
  \url{https://mikrotik.com/product/R52H}
\BIBentrySTDinterwordspacing

\bibitem{R5SHPn}
\BIBentryALTinterwordspacing
``Mikrotik {R5SHPn} {Data} {Sheet}.'' [Online]. Available:
  \url{https://routerboard.com/R5SHPn}
\BIBentrySTDinterwordspacing

\bibitem{physical_layer}
A.~Abdelgader and W.~Lenan, ``The {Physical} {Layer} of the {IEEE} 802.11p
  {WAVE} {Communication} {Standard}: {The} {Specifications} and {Challenges},''
  in \emph{Proc. of WCECS 2014}, vol.~2, Oct. 2014.

\bibitem{contention}
R.~Reinders \emph{et~al.}, ``Contention {Window} {Analysis} for {Beaconing} in
  {VANETs},'' in \emph{2011 7th International Wireless Communications and
  Mobile Computing Conference}, Jul. 2011, pp. 1481--1487.

\bibitem{rtp}
H.~Schulzrinne \emph{et~al.}, ``{RTP}: {A} {Transport} {Protocol} for
  {Real}-{Time} {Applications},'' United States, 2003.

\bibitem{receiver_sens}
A.~Paier, D.~Faetani, and C.~F. Mecklenbräuker, ``Performance {Evaluation} of
  {IEEE} 802.11p {Physical} {Layer} {Infrastructure}-to-{Vehicle}
  {Real}-{World} {Measurements},'' in \emph{2010 3rd International Symposium on
  Applied Sciences in Biomedical and Communication Technologies (ISABEL 2010)},
  Nov. 2010, pp. 1--5.

\bibitem{rician}
O.~Renaudin \emph{et~al.}, ``Wideband {MIMO} {Car}-to-{Car} {Radio} {Channel}
  {Measurements} at 5.3 {GHz},'' in \emph{2008 IEEE 68th Vehicular Technology
  Conference}, Sep. 2008, pp. 1--5.

\end{thebibliography}
\end{document}